\documentstyle[aas2pp4]{article}

\begin{document}

\title{An Optical Precursor to the Recent 
X-ray Outburst of the Black Hole Binary GRO J1655-40
}

\author{Jerome A. Orosz\altaffilmark{1}}
\affil{Department of Astronomy \& Astrophysics, The Pennsylvania State
University, 525 Davey Laboratory, University Park, PA  16802-6305 \\
orosz@astro.psu.edu}

\author{Ronald A. Remillard\altaffilmark{1}}
\affil{Center for Space Research, Massachusetts Institute of Technology,
Cambridge, MA 02139-4307 \\ rr@space.mit.edu}

\author{Charles D. Bailyn\altaffilmark{2}}
\affil{Department of Astronomy, Yale University, P.O. Box 208101,
New Haven, CT 06520-8101  \\ 
bailyn@astro.yale.edu}

\and

\author{Jeffrey E. McClintock}
\affil{Harvard-Smithsonian Center for Astrophysics, 60 Garden Street,
Cambridge,  MA  02138 \\
jem@cfa.harvard.edu}

\altaffiltext{1}{Visiting Astronomer at Cerro Tololo Inter-American 
Observatory 
(CTIO), which is operated by the Association of Universities for Research 
in Astronomy Inc., under contract with the National Science
Foundation.} 
\altaffiltext{2}{National Young Investigator}

\begin{abstract}
The All Sky Monitor on the {\em Rossi X-ray Timing Explorer} detected
an X-ray (2-12 keV) outburst from the black hole binary GRO J1655-40
beginning near April 25, 1996.  Optical photometry obtained April
20-24, 1996 shows a steady brightening of the source in $B$, $V$,
$R$, and $I$ beginning about six days before the start of the X-ray
outburst.  The onset of the optical brightening was earliest in $I$
and latest in $B$.  However, the rate of the optical brightening was
fastest in $B$ and slowest in $I$.  The order of the increases in the
different optical filters suggests that the event was an
``outside-in'' disturbance of the accretion disk.  The substantial
delay between the optical rise and the rise of the X-rays may provide
indirect support for the advection-dominated accretion flow model of
the inner regions of the accretion disk.
\end{abstract}


\section{Introduction}\label{intro}

GRO J1655-40 was discovered July 27, 1994 with the Burst and Transient
Source Experiment (BATSE) on board the {\em Compton Gamma Ray
Observatory} (Zhang et al.\ 1994\markcite{zh94}).  Unlike most ``X-ray
novae,'' GRO J1655-40 continued to have major outburst events in hard
X-rays long after its initial high-energy outburst (see Harmon et al.\
1995a\markcite{ha95a}).  There was an outburst event in late March
1995 (Wilson et al.\ 1995)\markcite{whzpf95} and another one starting
in late July 1995 (Harmon et al.\ 1995b\markcite{ha95b}).  The
relatively short recurrence times between the X-ray outbursts may be
due to the relatively large value of the average mass transfer rate
[$\dot{M}_2=3.4\times 10^{-9} \,M_{\sun}$~yr$^{-1}$, see Orosz \&
Bailyn 1997 (hereafter OB97\markcite{OB97})].  After the July/August
1995 hard X-ray outburst, the source apparently settled into true
X-ray quiescence.  From the period of late August 1995 to April 1996,
the source was not detected by BATSE (Robinson et al.\
1996\markcite{ro96}).  The {\em ASCA} X-ray observatory made several
pointed observations in late March 1996 and found the X-ray luminosity
(2-10 keV) of the source to be quite low: $L_x\approx 2\times
10^{32}$~ergs~s$^{-1}$ (see Robinson et al.\ 1996\markcite{ro96},
Y. Ueda private communication).  This extended period of X-ray
quiescence ended in late April 1996 when the all sky monitor (ASM) on
the {\em Rossi X-ray Timing Explorer} ({\em RXTE}) satellite detected
an increase in the 2 to 12 keV X-ray flux (Remillard et al.\
1996\markcite{rem96}; Levine et al.\ 1996).  The source was again
detected by BATSE starting in late May 1996 (Harmon et al.\ 1996).
GRO J1655-40 has remained a persistent X-ray source for several months
after the April outburst, unlike the case in 1994-1995 where there
were several shorter outburst events.

From optical observations made during late April and early May, 1995
Bailyn et al.\ (1995)\markcite{bomr95b} established the spectroscopic
period ($2\fd 601\pm 0\fd 027$) and the mass function ($3.16\pm
0.15\,M_{\sun}$) of the system, thereby establishing GRO J1655-40 as
one of the latest members of the class of objects often referred to as
``black hole X-ray transients.''  The $B$, $V$, $R$, and $I$ light
curves obtained February and March 1996 (during the X-ray quiescence)
by OB97\markcite{ob97} were dominated by ellipsoidal variations from
the secondary star.  The grazing eclipse of the accretion disk by the
star allowed OB97\markcite{ob97} to model the March 1996 light curves
to find precise values of the orbital inclination ($i=69.5\pm
0.08\deg$) and the mass ratio ($Q=2.99\pm 0.08$).  The model fits to
all four filters were excellent---the standard deviations of the
residuals were less than 0.02 magnitudes.  Furthermore, the $V$ and
$I$ light curves showed no noticeable change between February 1996 and
March 1996, indicating that GRO J1655-40 probably had reached its
minimum overall brightness level.  Here we report additional
photometry of GRO J1655-40 obtained late April 1996, starting
fortuitously close to the onset of the X-ray outburst.  Using the
March 1996 light curves to establish a ``quiescent'' brightness level
for each filter, we demonstrate below that the April photometry shows
clear evidence for an optical brightening well {\em before} the
brightening in the X-rays occurred.  We discuss below the optical and
X-ray observations and reductions, the April 1996 optical light curves
and the optical precursor, and the possible implications these
observations have for various models of the outburst cycles and
quiescent systems.

\section{Observations and Reductions}\label{obs}

\begin{deluxetable}{ccccc}
\tablewidth{0pt}
\tablecaption{Fitted Parameters for the April Light Curve Residuals}
\tablehead{
\colhead{filter}             &
\colhead{slope}      &
\colhead{coefficient of}  &
\multicolumn{2}{c}{{fitted time of initial rise}}
          \\ 
\colhead{} & \colhead{(mag/day)} & \colhead{correlation} &
\colhead{(HJD 2,449,000+)} & \colhead{(UT day in 1996)}  }
\startdata
$I$      &  $-0.0423\pm 0.0019$   &  $-0.96$  &  $1192.76\pm 0.29$ &
            April $19.25\pm 0.29$ \nl
$R$      &  $-0.0453\pm 0.0018$   &  $-0.97$  &  $1192.88\pm 0.26$ &
            April $19.37\pm 0.26$ \nl
$V$      &  $-0.0553\pm 0.0012$   &  $-0.99$  &  $1193.32\pm 0.15$ &
            April $19.82\pm 0.15$ \nl
$B$      &  $-0.0759\pm 0.0020$   &  $-0.99$  &  $1193.84\pm 0.18$ &
            April $20.34\pm 0.18$  \nl
X-ray\tablenotemark{a}    &  $20.44\pm 1.09$\tablenotemark{b} & $+0.97$  &
$1198.88\pm 0.78$  &     April $25.38\pm 0.78$  \nl
\enddata
\tablenotetext{a}{Determined from a fit to 23 points between HJD
2,450,199.09 and HJD 2,450,203.18.}
\tablenotetext{b}{The units are counts per second per day.}
\tablecomments{All errors shown are $1\sigma$.}
\label{fits}
\end{deluxetable}

\subsection{Optical Observations}

Photometry of the source was obtained April 20-24, 1996 with the CTIO
0.9 meter telescope, the Tek $2048\times 2046$ \#3 CCD, and standard
$B$, $V$, $R$, and $I$ filters.  The night of April 20 was photometric
and we observed a total of 59 Landolt (1992)\markcite{landolt92} stars
in five different fields, using the standard filters.  IRAF tasks were
used to process the images to remove the electronic bias and to
perform the flat-field corrections.  The programs DAOPHOT IIe and
DAOMASTER (Stetson 1987\markcite{st87}; Stetson, Davis, \& Crabtree
1991\markcite{sdc91}; Stetson 1992a\markcite{st92a},b) were used to
compute the photometric time series of GRO J1655-40 and several field
comparison stars.  The differences between the calibrated magnitudes
of nine stable field stars and their mean DAOPHOT instrumental
magnitudes were computed and the averaged differences were used to
correct the rest of the stars' instrumental magnitudes to the standard
system.  The uncertainty of the transformations between the DAOPHOT
instrumental magnitudes and the calibrated magnitudes are $0.056$ mag
in $B$, $0.006$ mag in $V$, $0.006$ mag in $R$, and $0.009$ mag in
$I$.  Note that these errors reflect the uncertainty in the zero
points of the magnitude scales---the internal errors of the
instrumental magnitudes within each filter are much smaller.  The
relatively large error of the $B$ transformation is because the color
term in $B$ is about ten times larger than the other three color
terms, resulting in more scatter in the differences between the
standard $B$ magnitudes and the instrumental $B$ magnitudes.  We find
that the photometric calibrations from the March and April data sets
are the same to 0.005 magnitudes.

\subsection{X-ray Observations}

A recent paper by Levine et al.\ (1996)\markcite{levine} describes the
ASM instrument on the {\em RXTE} and the data analysis procedures.
The ASM has been operating more or less continuously since February
21, 1996, providing roughly 5 to 10 scans of a given source per day.
GRO J1655-40 was not detected above the level of $\approx 12$~mCrab
before April 25, 1996 (Remillard et al.\ 1996\markcite{rem96}; Levine
et al.\ 1996\markcite{levine}).  After that, the intensities derived
from daily averages of the measurements were $71\pm 15$~mCrab on April
25, $420\pm 30$~mCrab on April 26, $586\pm 15$~mCrab on April 27,
$802\pm 9$~mCrab on April 28, and $1077\pm 30$~mCrab on April 29
(e.g.\ Remillard et al.\ 1996\markcite{rem96}).  The X-ray light
curve of GRO J1655-40 is given in Levine et al.\
(1996)\markcite{levine}.  We obtained the X-ray data from the public
archive maintained by the {\em RXTE} Guest Observer Facility.

\section{Optical Light Curves and the Optical Precursor}

\begin{figure*}[p]
\epsscale{1.7}
\plotone{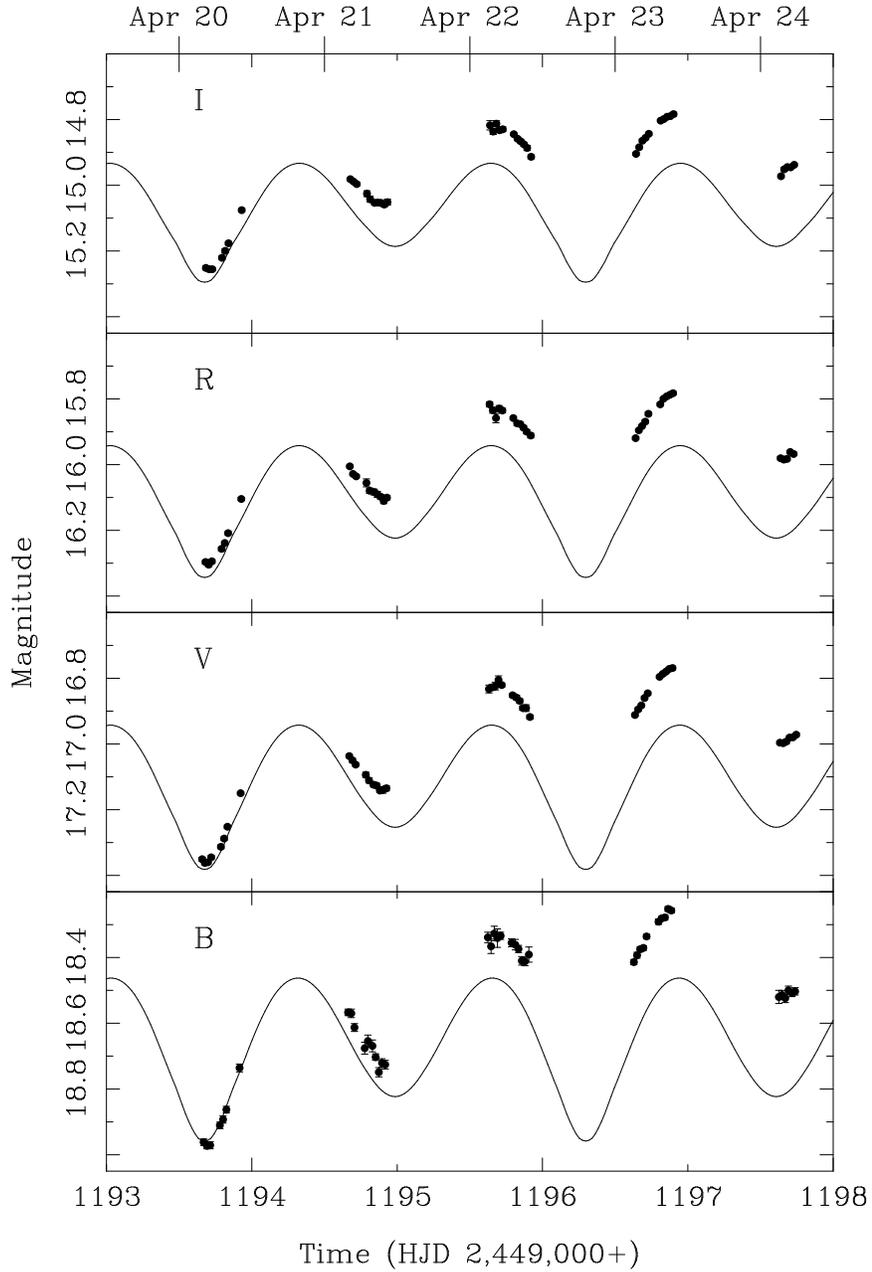}
\caption{The  observed light curves
of GRO J1655-40 from April 20-24, 1996 (points) and the models which fit
the light curves from March 18-25, 1996 (solid lines).  The same magnitude 
range (0.85 mag) is shown for all four panels, demonstrating the 
notable decrease in the light curve amplitudes from $B$ to $I$.  
There is a clear deviation of the data from the models
in all four filters.}
\label{newfig1}
\end{figure*}

\begin{figure*}[p]
\epsscale{1.7}
\plotone{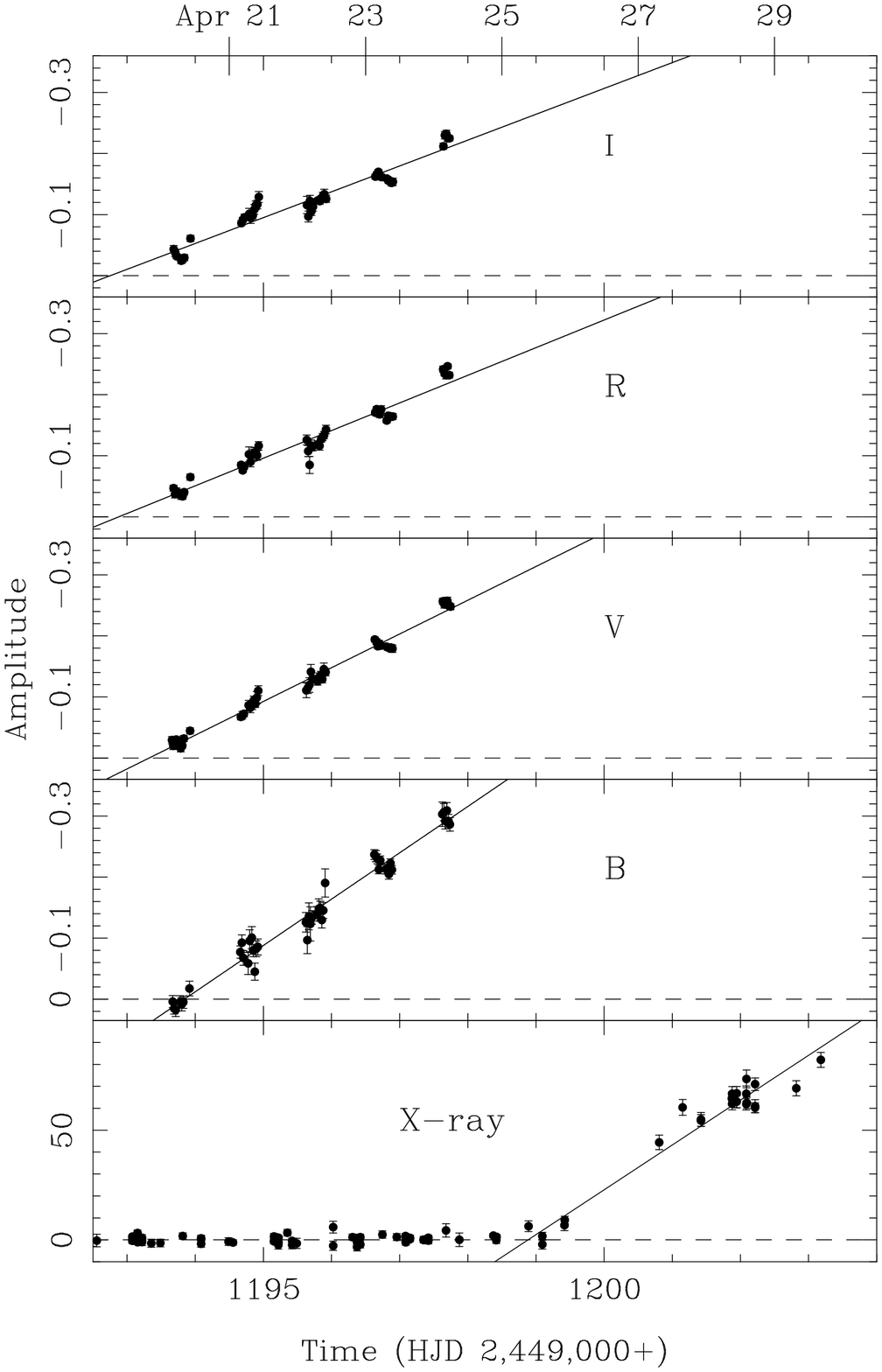}
\caption{The residuals in the sense of the April photometry data {\em minus}
the March models
(in magnitudes) and the best-fitting lines are shown in the upper 4 panels.
The bottom panel shows the X-ray intensity in counts per second (2 to 12 keV)
as a function of time and the best
fitting straight line to the 23 points between
April 25.5 and 29.5.
The measurements from the individual dwells and their
uncertainties are shown (see Levine et al.\ 1996).  
The intensity of the Crab Nebula is about 75 counts per second.
See Table \protect\ref{fits} for the linear fit parameters.
}
\label{newfig2}
\end{figure*}

The eclipsing light curve code used to model the March 1996 photometry
is described in some detail by OB97\markcite{ob97}.  Briefly, the code
uses full Roche geometry to account for the distorted secondary star.
Light from a circular accretion disk and the effects of mutual
eclipses are also accounted for.  The code computes the observed flux
as a function of the orbital phase, so the $BVRI$ light curves from
March 1996 were each ``folded'' using the spectroscopic ephemeris
given in OB97\markcite{ob97} and fitted simultaneously.  The best
fitting model consists of four different curves giving the $B$, $V$,
$R$, and $I$ magnitudes as a function of the photometric phase.  These
four curves were ``unfolded'' using the spectroscopic ephemeris,
yielding model curves which give the magnitude as a function of time.

The orbital phase during our April observations is determined
precisely (to within 0.008 of an orbital cycle) by the spectroscopic
ephemeris given in OB97\markcite{ob97}.  Similarly, the shapes and
amplitudes of the model light curves for the quiescent state, which
are shown in Figure \ref{newfig1}, are precisely specified in OB97.
Figure \ref{newfig1} shows the April photometry plotted with these
model curves.  There is a clear systematic deviation of the
observations from the model in all four filters.  Figure \ref{newfig2}
shows the residuals in the sense of the April data {\em minus} the
March model (propagated forward in time to April).  The systematic
deviation from the model in all four filters is now quite clear: the
source increased in brightness starting near April 20.  In each case,
the residuals on a magnitude scale decrease linearly with time (see
Table \ref{fits} for the parameters of the linear fits to the
residuals).  An inspection of Figure \ref{newfig2} and Table
\ref{fits} shows some interesting trends.  First, the rate of the
brightness increase is the largest in $B$ ($0.0759\pm 0.0020$ mag/day)
and the smallest in $I$ ($0.0423\pm 0.0019$ mag/day).  However, the
fitted time of the initial rise is the earliest in $I$ (April
$19.25\pm 0.29$ UT) and the latest in $B$ (April $20.34\pm 0.18$ UT).

We took the 23 X-ray observations between HJD 2,450,199.09 and HJD
2,450,203.18 (roughly April 25.5 to 29.5, 1996) and fit a straight
line to estimate the time of the initial X-ray rise.  The results of
the fit are shown in Table \ref{fits} and in Figure \ref{newfig2}. The
initial rise in the X-ray occurred April $25.38\pm 0.78$, just over
six days after the initial rise in the $I$ filter (April $19.25\pm
0.29$).

\section{Discussion}

The quiescent state of the black hole transients has been the subject
of intense study.  Recently, Narayan, McClintock, \& Yi (1996,
hereafter NMY96) presented their advection-dominated accretion flow
(ADAF) model for the quiescent black hole binary systems.  In the
NMY96 model, there is an outer region of the disk at radii larger than
several thousand Schwarzschild radii consisting of a thin disk and a
zone interior to this where the accretion flow is a hot,
optically-thin and quasi-spherical ADAF.  In this ADAF region, the
energy released by viscous dissipation is largely swept along with the
flow and enters the black hole before it can be radiated.
Consequently, the ADAF model is $\approx 100-1000$ times less
efficient at producing X-rays than the standard disk model, a fact
which can explain the quiescent spectra of A0620-00 and V404 Cyg
(NMY96).

The two classes of models used to explain the cause of the outbursts
of the dwarf novae (i.e.\ accreting white dwarf systems), the disk
instability model (DIM; Cannizzo, Chen, \& Livio 1995\markcite{ccl95})
and the mass-transfer instability model (MTI; Hameury, King, \&
Lasota, 1986\markcite{hkl86}, 1987\markcite{hkl87},
1988\markcite{hkl88}, 1990\markcite{hkl90}), have in the past been
applied to the soft X-ray transients.  However, the MTI model as
applied to the soft X-ray transients (especially those with periods
less than about 9 hours) has recently fallen out of favor (Lasota
1996a\markcite{las96a},b\markcite{las96b}; Tanaka \& Shibazaki
1996\markcite{ts96}).  Also, the ``standard'' DIM, where the inner
disk extends all the way down to the innermost stable orbit (about
three times the Schwarzschild radius), has some difficulties when
applied to the black hole X-ray transients (e.g.\ Lasota, Narayan, \&
Yi 1996\markcite{lny96}; Lasota 1996a\markcite{las96a},
1996b\markcite{las96b}; Tanaka \& Shibazaki 1996\markcite{ts96}).
Therefore both ``standard'' theories need some modification when
applied to the black hole X-ray transients.

Lasota, Narayan, \& Yi (1996)\markcite{lny96} developed a model for
the outbursts in the black hole X-ray transients based on the NMY96
model.  In particular, Lasota et al.\ (1996) show that the outburst
mechanism in the X-ray transients must be associated with an
instability in the cold outer disk of the NMY96\markcite{nmy96} model
(i.e.\ the advection dominated region cannot be responsible for the
triggering of the outbursts).  They go on to show that the instability
could be a pure thermal accretion disk instability (moving either
inside-out or outside-in) or an instability caused by enhanced mass
transfer from the secondary where a small amount of additional matter
causes a marginally stable disk to become unstable.

Our photometry results strongly suggest that the instability that
created the April 1996 outburst started in the outer regions of the
disk, and propagated inward (an ``outside-in'' event).  The
temperature profile of the accretion disk in the OB97\markcite{ob97}
model is parameterized as $T(r)=T_{\rm disk}(r/r_{\rm disk})^{\xi}$.
The best value of $\xi$ was negative ($-0.12\pm 0.01$), indicating
that the disk gets hotter as one moves inward.  As a result, the inner
regions of the disk are bluer than the outer regions of the disk.
From Figure \ref{newfig2} and Table \ref{fits} we see that the optical
brightening started first in the $I$ filter and last in the $B$
filter.  Since the optical brightening in GRO J1655-40 appears first
in the $I$ band, the only region where this would appear as a ``warm''
zone is in the outer disk. Thus, it is reasonable to conclude that the
disturbance in the accretion disk that gave rise to the increase in
the optical flux was indeed an outside-in event.

Since either inside-out or outside-in transition waves are allowed in
the Lasota et al.\ (1996) model, our demonstration that an outside-in
event triggered the April 1996 outburst does not seem to constrain
this outburst model.  Paradoxically, our observation of the optical
outburst might actually provide evidence for the presence of the ADAF
region in the NMY96 model of the quiescent system.  Consider the
$\approx 6$ day delay between the onset of the optical brightening and
the brightening in the X-rays.  This is analogous to the situation in
certain dwarf novae systems where the rise in the optical flux
precedes the rise in the UV flux by roughly one day (see Meyer \&
Meyer-Hofmeister 1994\markcite{mmh94}).  In the standard DIM the
transition wave should travel from the optical-emitting outer edge of
the disk to the UV-emitting inner edge of the disk in only a few
hours.  Meyer \& Meyer-Hofmeister (1994)\markcite{mmh94} suggested
that a ``hole'' in the inner disk could explain the observed
optical-UV lag since the hole would take time to fill up before the
hot UV-emitting region can be established (see Cannizzo
1993\markcite{can93} for other possible solutions).  In the case of
GRO J1655-40, the ADAF region of the NMY96 model would act as the hole
in the middle of the disk since the standard thin disk is truncated at
the edge of the ADAF region.  The fact that GRO J1655-40 is a much
larger system than a typical dwarf nova 
(i.e. the orbital separation of GRO J1655-40 is $a=16.77\pm 0.19\,R_{\sun}$
[OB97] compared to a separation of $a=3.29\pm 0.29\,R_{\sun}$ for
SS Cyg [derived from parameters given in Hessman et al.\
1984]) 
explains why the the
optical--X-ray lag in GRO J1655-40 is longer than the optical-UV lag
in the dwarf novae.  Detailed computations would be required to see if
the $\approx 6$ day delay between the optical and X-ray can be
explained by the NMY96\markcite{nmy96} model---such computations are
well beyond the scope of this paper.

\section{Summary}

It appears that GRO J1655-40 was in a true quiescent state before the
April 1996 outburst because of the low X-ray flux observed in March
1996 and the nature of the optical light curves from February and
March 1996.  We observed an increase in the optical flux about six
days before the start of the X-ray outburst (which started near April
25.4).  The character of the optical outburst depends on the color:
the rate of increase was the fastest in $B$ and the slowest in $I$
whereas the brightening began first in $I$ and last in $B$.  The order
in which the optical brightening started (from the redder colors to
the bluer colors) suggests that the disturbance in the disk was an
``outside-in'' one.  The substantial delay between the optical rise
and the rise of the X-rays may provide indirect support for the ADAF
model of the quiescent black hole binaries.  Further detailed
computations are needed to confirm this.

\acknowledgements

We are grateful to the staff of CTIO for the excellent support, in
particular Srs.\ Maurico Navarrete, Edgardo Cosgrove, Maurico
Fernandez, \& Luis Gonzalez. We thank Jean-Pierre Lasota for the
comments on an earlier draft of this paper. Partial financial support
for this work was provided by the National Science Foundation through
a National Young Investigator grant to C. Bailyn and by the
Smithsonian Institution Scholarly Studies Program to J.  McClintock.


\begin{references}

\reference{bomr95b}
Bailyn, C. D., Orosz, J. A., McClintock, J. E., \& Remillard, R. A.
1995, Nature, 378, 157

\reference{can93}
Cannizzo, J. K. 1993, in ``Accretion Disks in Compact Stellar Systems'',
ed. J. C. Wheeler, World Scientific Publishing Co., Singapore

\reference{ccl95}
Cannizzo, J. K., Chen, W., \& Livio, M. 1995, \apj, 454, 880

\reference{hkl86}
Hameury, J. M., King, A. R., \& Lasota, J. P. 1986, \aap, 161, 71

\reference{hkl87}
Hameury, J. M., King, A. R., \& Lasota, J. P. 1987, \aap, 171, 140

\reference{hkl88}
Hameury, J. M., King, A. R., \& Lasota, J. P. 1988, \aap, 192, 187

\reference{hkl90}
Hameury, J. M., King, A. R., \& Lasota, J. P. 1990, \apj, 353, 585

\reference{ha95a}
Harmon, B. A., et al.\ 1995a, Nature, 374, 703

\reference{ha95b}
Harmon, B. A., et al.\ 1995b, IAU Circular \#6205

\reference{ha96}
Harmon, B. A., et al.\ 1996, IAU Circular \#6436

\reference{hess84}
Hessman, F. V., Robinson, E. L., Nather, R. E., \& Zhang, E.- H.
1984, \apj, 286, 747


\reference{landolt92}
Landolt, A. U. 1992,\aj, 104, 340

\reference{las96a}
Lasota, J. P. 1996a, in ``Compact Stars in Binaries'', Proc.\
of IAU Symp.\ 165, eds.\ J. van Paradijs, E. P. J. van den Heuvel,
\& E. Kuulkers, Kluwer Academic Press, Dordrecht.

\reference{las96b}
Lasota, J. P. 1996b, in ``Accretion
Phenomena and Related Outflows'', Proc.\ IAU Coll.\
163, eds.\ D. Wickramasinghe, L. Ferrario, \&
G. Bicknell, in press

\reference{lny96}
Lasota, J. P., Narayan, R., \& Yi, I. 1996, \aap, 314, 813

\reference{levine}
Levine, A. M. et al.\ 1996, \apj, 469, L33

\reference{mmh94}
Meyer, F., \& Meyer-Hofmeister, E. 1994, \aap, 288, 175

\reference{nmy96}
Narayan, R., McClintock, J. E., \& Yi, I. 1996, \apj, 451, 821
(NMY96)

\reference{ob97}
Orosz, J. A., \& Bailyn 1997, \apj, in press, to appear March 10, 1997 (OB97)

\reference{rem96}
Remillard, R. A., Bradt, H., Cui, W., Levine, A., Morgan, E., Shirey,
B., \& Smith, D. 1996, IAU Circular \#6393

\reference{ro96}
Robinson, C., et al.\ 1996, \apj, submitted

\reference{st87} 
Stetson, P. B.\ 1987, \pasp, 99, 191
 
\reference{sdc91}
Stetson, P. B., Davis, L. E., \& Crabtree, D. R.\ 1991, in
``CCDs in Astronomy,'' ed.\ G. Jacoby, ASP Conference Series, Volume
8, page 282
 
\reference{st92a} 
Stetson, P. B.\ 1992a, in ``Astronomical Data Analysis Software and 
Systems I,'' eds.\ D. M. Worrall, C. Biemesderfer, \& J. Barnes, ASP 
Conference Series, Volume 25, page 297
 
\reference{st92b} 
Stetson, P. B.\ 1992b, in ``Stellar Photometry---Current Techniques and 
Future Developments,'' IAU Coll.\ 136, eds.\ C. J. Butler, \& I. Elliot, 
Cambridge University Press, Cambridge, England,
page 291

\reference{ts96}
Tanaka, Y., \& Shibazaki, N. 1996, \araa, 34, 607

\reference{whzpf95}
Wilson, C. A., Harmon, B. A., Zhang, S. N., Paciesas, W. S.,
\& Fishman, G. J. 1995, IAU Circular \#6152

\reference{zh94}
Zhang, S. N., et al.\ 1994, IAU Circular \#6046

\end{references}
\end{document}